\documentclass[journal=jctcce,manuscript=article]{achemso}

\usepackage{amsmath}
\usepackage{color}
\usepackage{graphicx}
\newcommand{\Hef}{{}^{4\!}\mbox{He}}

\newcommand{\apHe}{\bar{p}\,\mbox{H}\mbox{e}^{+}}
\newcommand{\apHet}{\bar{p}{}^{\,3\!}\mbox{H}\mbox{e}^{+}}
\newcommand{\apHef}{\bar{p}{}^{\,4\!}\mbox{H}\mbox{e}^{+}}
\author{Hubert J. Jóźwiak}
\affiliation[NCU University]
{Institute of Physics, Faculty of Physics, Astronomy and Informatics, Nicolaus Copernicus University in Torun, Grudziadzka 5, 87-100 Torun, Poland}
\email{hubert.jozwiak@umk.pl}
\author{Dimitar Bakalov}
\affiliation[INRNE Sofia]
{Institute for Nuclear Research and Nuclear Energy, Bulgarian Academy of Sciences, Sofia, Bulgaria}
\author{Michał Przybytek}
\affiliation[Warsaw University]
{University of Warsaw, Faculty of Chemistry, Pasteura 1, 02-093 Warsaw, Poland}
\author{Michail Stoilov}
\affiliation[INRNE Sofia]
{Institute for Nuclear Research and Nuclear Energy, Bulgarian Academy of Sciences, Sofia, Bulgaria}
\author{Piotr Wcisło}
\affiliation[NCU University]
{Institute of Physics, Faculty of Physics, Astronomy and Informatics, Nicolaus Copernicus University in Torun, Grudziadzka 5, 87-100 Torun, Poland}
\title[An \textsf{achemso} demo]
  {Quantum calculation of the collision-induced line-shape effects in antiprotonic helium and the new accurate ab initio $\apHe$--He potential energy surface}

\abbreviations{IR,NMR,UV}

\begin{document}

%%%%%%%%%%%%%%%%%%%%%%%%%%%%%%%%%%%%%%%%%%%%%%%%%%%%%%%%%%%%%%%%%%%%%
%% The "tocentry" environment can be used to create an entry for the
%% graphical table of contents. It is given here as some journals
%% require that it is printed as part of the abstract page. It will
%% be automatically moved as appropriate.
%%%%%%%%%%%%%%%%%%%%%%%%%%%%%%%%%%%%%%%%%%%%%%%%%%%%%%%%%%%%%%%%%%%%%
%%%%%%%%%%%%%%%%%%%%%%%%%%%%%%%%%%%%%%%%%%%%%%%%%%%%%%%%%%%%%%%%%%%%%
%% The abstract environment will automatically gobble the contents
%% if an abstract is not used by the target journal.
%%%%%%%%%%%%%%%%%%%%%%%%%%%%%%%%%%%%%%%%%%%%%%%%%%%%%%%%%%%%%%%%%%%%%
\begin{abstract}
  We present the first fully \emph{ab initio} calculations of collision-induced broadening and shift of spectral lines in antiprotonic helium ($\apHe$) perturbed by atomic helium. To overcome critical limitations of previous studies, we construct a new highly accurate potential energy surface (PES) that spans a wide range of $\apHe$--He geometries relevant to all metastable states of the exotic helium atom. Rigorous quantum scattering calculations performed using the new PES yield scattering $S$-matrices from which we extract pressure broadening and shift coefficients for 50 transitions in antiprotonic helium-4 ($\apHef$). This dataset provides the first rigorous benchmark for earlier semiclassical calculations and establishes a robust theoretical reference for high-precision spectroscopy of antiprotonic helium, which is used to test the fundamental charge, parity, and time reversal (CPT) symmetry. The results extend to temperatures relevant to non-gaseous phases of helium, supporting a new class of precision measurements. This study introduces methodological framework for future investigations of other exotic systems, such as pionic or kaonic helium atoms, enabling the development of reference data for high-precision spectroscopy of these species---an essential component for improving the determination of the pion and kaon masses.
\end{abstract}

%%%%%%%%%%%%%%%%%%%%%%%%%%%%%%%%%%%%%%%%%%%%%%%%%%%%%%%%%%%%%%%%%%%%%
%% Start the main part of the manuscript here.
%%%%%%%%%%%%%%%%%%%%%%%%%%%%%%%%%%%%%%%%%%%%%%%%%%%%%%%%%%%%%%%%%%%%%

\section{Introduction}

Exotic antiprotonic helium atoms are three-body bound systems consisting of a helium nucleus, an antiproton, and an electron. 
These exotic systems combine features of a diatomic molecule (two heavy charged particles bound by an orbiting electron) and of a two-electron atom. 
Antiprotonic helium atoms $\apHe$ are formed when negatively charged antiprotons are slowed down in helium gas and captured by the Coulomb field of helium nuclei. 
Antiprotons are initially captured at highly excited orbitals, predominantly in short-living excited states that de-excite via fast Auger transitions within nanoseconds down to states in which the large overlap of the antiproton with the nucleus causes immediate annihilation. 
A small fraction of antiprotons of the order of 3\%, however, is captured in near-circular metastable states with the principal ($n$) and orbital ($l$) quantum numbers, $n\sim38$, $l\sim n-1$, and lifetime of the order of microseconds.
Here, the hydrogen-like quantum numbers $n,l$ refer to the antiproton orbital in the field of He$^+$ while the electron of $\apHe$ is assumed to occupy the lowest-energy electronic orbital.
The long lifetime of these metastable states is determined by the lower rate of de-excitation through slow radiative transitions since Auger transitions are suppressed. 
This phenomenon was first pointed out by Condo\cite{Condo_1964} in his study of the decay of negative pions in helium.

The existence of long-living metastable states offers unique opportunities for high-precision laser spectroscopy of antiprotonic helium and opens room for measurements of the mass and dipole magnetic moment of antiprotons and independent tests of the fundamental CPT invariance.\cite{Hayano_2007}
In the first generation of $\apHe$ experiments performed predominantly in the 1990s at the CERN Low Energy Antiproton Ring (LEAR) and thoroughly reviewed in Refs.~\cite{Eades_1999,Yamazaki_2002}, the antiprotons were stopped in helium gas target of atomic number density of the order of $10^{21}$~cm$^{-3}$. 
At such densities, the pressure broadening and shift of the laser-stimulated $E1$-transition spectral lines is comparable with the leading order relativistic and QED effects and turns out to be the main systematic effect that limits the experimental accuracy. 
To compare the measured transition frequencies with theoretical calculations \cite{Korobov_1996}, the experimental results were extrapolated to zero helium gas density using a semiclassical method for the evaluation of the pressure effects and \emph{ab initio} potential energy surface (PES)\cite{Bakalov_2000}. 
This led to the first experimental determination of the dipole magnetic moment of antiproton \cite{Pask_2009}.

The commissioning of the CERN Antiproton Decelerator (AD) in 2000, of the Radiofrequency Quadrupole Decelerator apparatus (RFQD), and, later, of the Extra Low ENergy Antiproton ring (ELENA) in 2017, enabled a new generation of higher accuracy antiprotonic helium spectroscopy experiments. 
The low energy of the incident antiprotons (5.3 MeV for AD, 100 keV for ELENA, and $\sim60$ keV for RFQD) made it possible to stop them in a helium gas target at density as low as $\sim10^{16}$ cm$^{-3}$, thus strongly suppressing the expected density broadening and shift of the transition frequencies to sub-MHz levels \cite{Hayano_2008} and allowing to boost the fractional accuracy of the measurements to the ppb level \cite{Hori_2016}. 
In this way, the most accurate value of the antiproton-to-electron mass ratio was obtained \cite{Hori_2016}. 
A few years earlier the experimental value of the antiproton-to-electron mass ratio had been obtained by two-photon spectroscopy with a slightly larger uncertainty in a helium target at density $\sim10^{18}$ cm$^{-3}$, the main systematic error coming from the density shift \cite{Hori_2011}.

The family of exotic helium atoms extends beyond $\apHe$. 
The pionic helium atom has been shown to possess long-lived metastable states \cite{Hori_2020} that open perspectives for the measurement of the negative pion-to-electron mass ratio with high accuracy. 
However, this experimental approach is expected to face the problem of large pressure broadening and shift of the spectral lines, of the order of 100 GHz at the envisaged superfluid helium densities $\sim2 \times 10^{22}$~cm$^{-3}$\cite{Hori_2022}. 
Recent theoretical papers investigated the structure of kaonic helium atoms \cite{Korobov_2024}. Hypothetical kaonic helium spectroscopy could potentially be used to determine the negative kaon-to-electron mass ratio, offering the possibility of refining the best existing measurements, which achieve a relative precision of $10^{-5}$.\cite{Workman_2022_review}

This brief overview shows that precision exotic helium atom spectroscopy holds immense potential for accurate determination of the masses of exotic particles and testing the fundamental CPT symmetry. At the same time, one of the leading sources of systematic errors in such experiments arises from density (pressure) shifts and the broadening of the spectral lines.

Early studies at LEAR\cite{Torii_1999} and later at AD\cite{Hori_2001,Hori_2016}, as well as the two-photon spectroscopy experiment \cite{Hori_2011} sought to characterize the pressure effects experimentally by measuring transition frequencies at different helium densities and extrapolating to zero density assuming linear dependence of the resonance frequency on density and neglecting temperature dependence. While this approach enabled sub-ppm precision in individual frequency measurements, the extracted density shift coefficients had uncertainties of $\sim$10\% or higher \cite{Hori_2001,Hori_2016}.

The first theoretical evaluation of pressure broadening and shift in antiprotonic helium was reported in Ref.~\cite{Bakalov_2000}. 
The $\apHe$--He PES was evaluated using the symmetry-adapted perturbation theory (SAPT) \cite{Jeziorski_1994,Patkowski_2020} in the symmetrized Rayleigh-Schr\"{o}dinger (SRS)\cite{Jeziorski_1978} formulation. 
A brief description of this PES is provided in Section~\ref{sec:PES}. 
Pressure broadening and shift coefficients for a set of laser-stimulated $E1$ transitions in metastable $\bar{p}$He states were then evaluated using the semiclassical Anderson's method \cite{Anderson_1952}, in which the relative motion of helium atoms was treated classically. 
While the calculated pressure broadenings and shifts agreed with the available experimental data, later experiments revealed transitions for which theoretical predictions significantly diverged from the experiment.\cite{Yamazaki_2002,Hori_2001} 
The same PES (after appropriate coordinate transformation) was used in the evaluation of the density effects in pionic helium \cite{Obreshkov_2016} by applying in parallel Anderson's semiclassical method and Baranger's quantum method \cite{Baranger_1958a,Baranger_1958b}; the obtained results were in reasonable agreement with each other. 

Recently, the calculation of a new PES for the exotic-helium--helium pair was reported. 
To our knowledge, it has only been applied to collision-induced transitions between hyperfine states of antiprotonic helium and the collisional quenching rates of pionic helium \cite{Bibikov_2020,Bibikov_2023}. 
Other studies explored alternative approximations, such as estimating pressure broadening and shift of pionic helium lines using only the leading $1/R^{6}$ term in the long-range part of the $\pi$He--He interaction\cite{Bai_2022, Korobov_2015}. 
Additionally, density effects in antiprotonic helium spectra in liquid \cite{Adamczak_2013} or solid \cite{Adamczak_2014} helium targets were evaluated using combinations of the \emph{ab initio} PES with phenomenological interaction potentials.

Nevertheless, several critical limitations persist in previous evaluations of the pressure effects in exotic helium spectroscopy. 
First, the PES used in earlier calculations was constructed for a limited range of $\apHe$--He geometry configurations, optimized for only a subset of metastable states. 
Additionally, the sparse grid of computed interaction energies led to interpolation challenges, where different smooth fits diverged significantly outside the original grid, impacting the reliability of pressure broadening and shift predictions. 
Finally, under the specific experimental conditions (helium gas target density up to $\sim10^{21}$ cm$^{-3}$ and temperature of 5 K or lower) the validity of the semi-classical approximation, in which the inter-atomic dynamics is treated classically, is questionable, potentially leading to systematic errors in predicted density effects. 

In this work, we address these challenges by performing the first fully \emph{ab initio} calculations of the collisional perturbation of the antiprotonic helium lines using the new, state-of-the-art $\apHef$--$\Hef$ potential energy surface. 
Compared to the previous PESs \cite{Bakalov_2000}, the number of \emph{ab initio} points has been increased by almost two orders of magnitude, spanning the range of values relevant for all metastable states of antiprotonic and pionic helium. 
The PES has been used to perform quantum scattering calculations in the $\apHef$--$\Hef$ system. 
The resulting scattering $S$-matrix yielded pressure broadening and shift coefficients for 50 $E$1 transitions in $\apHef$ across a broad range of helium gas temperatures. 
This dataset establishes a new benchmark for modeling density effects in exotic helium spectroscopy and provides a crucial reference for future high-precision experiments. 
Additionally, this work paves the way for extending pressure broadening and shift studies to other exotic helium atoms, such as pionic and kaonic helium, which could be used in future spectroscopic studies of these systems, aimed at refining the best determinations of the pion- and kaon-to-electron mass ratios.

\section{\label{sec:PES} Ab initio potential energy surface for $\apHef$--$\Hef$}

\subsection{Geometry specification and grid points}

\begin{figure}[!t]
\includegraphics[width=0.95\textwidth]{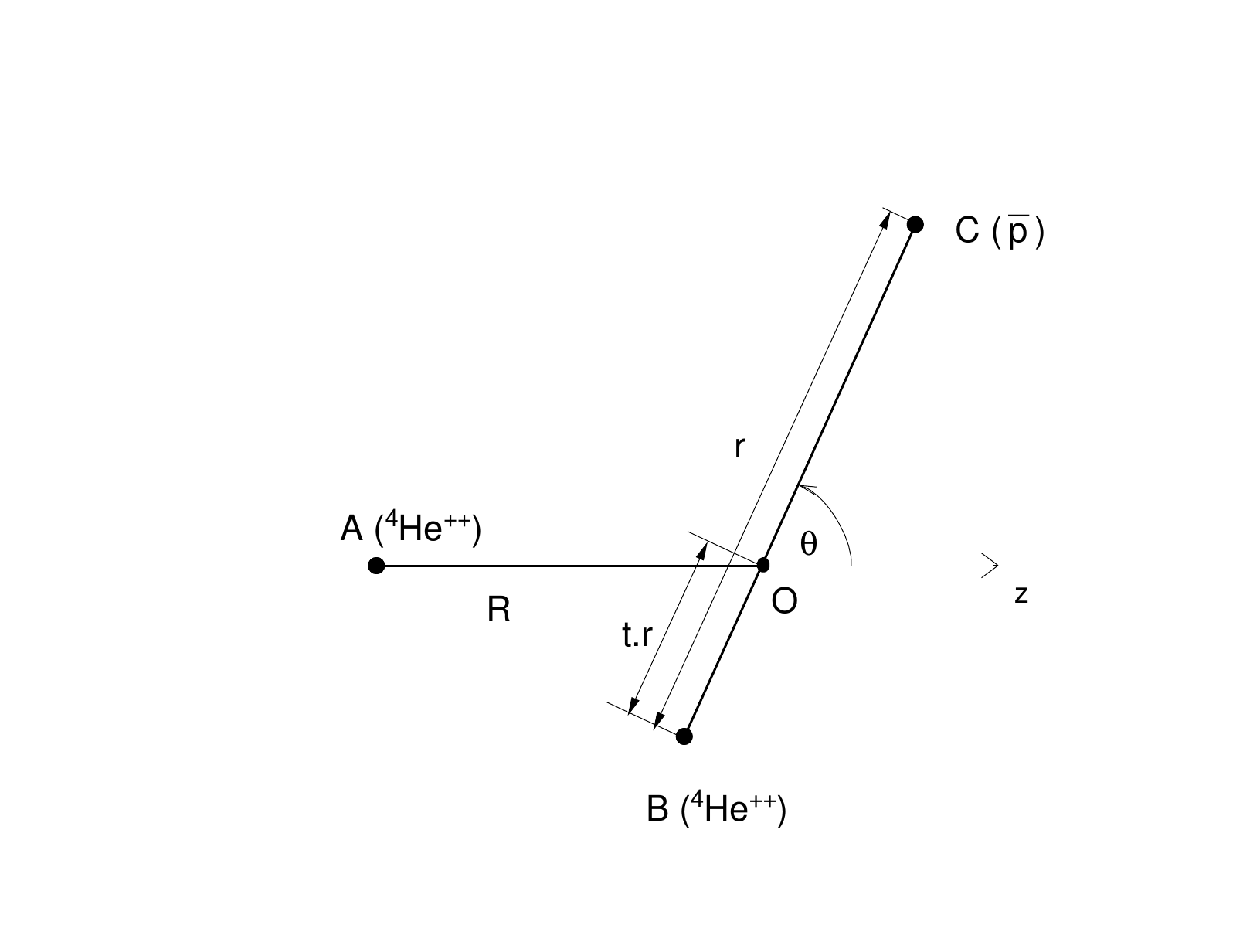}
\caption{Geometry of the $\apHef$--$\Hef$ system (see the text for details).}
\label{fig:geometry}
\end{figure}

Using the Born--Oppenheimer approximation we can separate the motion of electrons and the motion of heavy particles in a system consisting of exotic and ordinary helium atoms.
Let us denote the nucleus of the ordinary helium atom by A, the helium nucleus and the heavy negatively charged particle in the exotic helium atom by B and C, respectively, and the center of mass of B and C by O.
The electronic interaction energy in the system may then be parametrized with three Jacobi coordinates: the length $R$ of vector $\mathbf{R}$ pointing from A to O, the length $r$ of vector $\mathbf{r}$ pointing from B to C, and the angle $\theta$ between the vectors $\mathbf{R}$ and $\mathbf{r}$, see Fig.~\ref{fig:geometry}. 
The distance between the center O and the helium nucleus B, $r_\mathrm{OB}$, is determined by the masses of B and C: $r_\mathrm{OB}=tr$, where $t$ is defined as
\begin{equation}\label{eq:t}
t=m_\mathrm{C}/(m_\mathrm{B}+m_\mathrm{C})
\end{equation}
and varies with the type of exotic helium atom. For antiprotonic helium-4, considered here, this mass ratio calculated using the alpha particle mass $m_\mathrm{B}=m_\alpha=7294.29954171\,m_e$ and the proton mass $m_\mathrm{C}=m_p=1836.152673426\,m_e$ from 2022 CODATA recommended values \cite{CODATA22} is $t=0.201102$.

The previous potential energy surface for the $\apHef$--$\Hef$ system~\cite{Bakalov_2000} was evaluated for a set of 375 geometries. 
The grid points were selected to account for the average inter-particle distances in the metastable states of $\apHef$ that were being studied experimentally at that time. 
In particular, the $r$-values range was centered at 0.8~$a_{0}$, which corresponds to the average r.m.s. separation in states with $l\sim35$, in the center of the metastability domain. 
For lower or higher excited states, however, the average separation approaches the boundary of the $r$-range used in Ref.~\cite{Bakalov_2000}. 
Moreover, with the 375 points, the grid of $R$, $r$, and $\theta$ values was too sparse, leading to ill-defined fits to the \emph{ab initio} points. 
Different fits with similar values of $\chi^2$ could approximate the grid values well, but their extrapolations outside the grid differed qualitatively, leading to conflicting pressure broadening and shift coefficients for transitions between lower and higher excited states.

In this work, we provide a new potential energy surface for $\apHef$--$\Hef$ that addresses these challenges. 
The interaction energy for a pair of ordinary and antiprotonic helium-4 atoms is evaluated for the $45\times31\times19=26505$ combinations $(R_i,r_i,\theta_i)$, $i=1,\ldots,26505$, increasing the number of grid points by almost two orders of magnitude compared to the previous PES.\cite{Bakalov_2000} 
The parameters span the following ranges:
\begin{itemize} 
\item $R_i$ (45 values): $1.5-3.0\,a_{0}$ (step $0.5\,a_{0}$), $3.0-8.0\,a_{0}$ (step $0.25\,a_{0}$), $8.0-12.0\,a_{0}$ (step $0.5\,a_{0}$), $12.0-20.0\,a_{0}$ (step $1.0\,a_{0}$), and $20.0-30.0\,a_{0}$ (step $2.0\,a_{0}$); 
\item $r_{i}$ (31 values): $0.15-1.5\,a_{0}$ (step $0.05\,a_{0}$) and $1.5-1.8\,a_{0}$ (step $0.1\,a_{0}$); 
\item $\theta_{i}$ (19 angles): evenly spaced from $0^{\circ}$ to $180^{\circ}$ with $10^{\circ}$ increment. 
\end{itemize}

\subsection{Ab initio calculations}

The many-electron wave functions are represented using the full configuration interaction (FCI) approach~\cite{CI_method_1999} as an expansion in a set of Slater determinants constructed from Hartree--Fock spinorbitals, which are in turn expanded in a set of fixed one-electron basis functions.
The energies and wave functions of the considered many-electron states are obtained by the direct diagonalization of the Hamiltonian matrix calculated using all possible Slater determinants with a well-defined spatial symmetry and spin projection: $M_S=\frac12$ for the whole three-electron interacting system and the one-electron exotic helium atom, and $M_S=0$ for the two-electron ordinary helium atom.
To expand the spinorbitals, we use a family of doubly-augmented correlation-consistent polarized-valence Gaussian basis sets d$X$Z developed in Ref.~\cite{cencek_2012}, where the cardinal number $X$ ranges from $X=2$ to $X=7$, and the largest angular momentum quantum number of functions included in a given basis set is $l_\mathrm{max}=X-1$.
The one-electron basis functions are centered only on both helium nuclei, and the antiproton is treated as a singly negative point charge with no functions attached to it.
The FCI calculations were performed using the \textsc{Hector} program~\cite{hector}, while the Hartree--Fock orbitals and necessary one- and two-electron integrals were generated using the \textsc{Dalton} 2.0 package~\cite{Dalton2020,dalton2}. Due to the high computational demands of the FCI method, calculations with the largest basis set (d7Z) were performed on a smaller grid of 70 points and used only to analyze the basis set convergence pattern.

We calculate the interaction energy at any given geometry using the supermolecular approach,
\begin{equation}\label{eq:defVint}
V(R,r,\theta) = E_\mathrm{ABC}(R,r,\theta) 
- E_\mathrm{A} - E_\mathrm{BC}(r),
\end{equation}
where $E_\mathrm{ABC}(R,r,\theta)$ is the energy of the whole system (dimer) and $E_\mathrm{A}$ and $E_\mathrm{BC}(r)$ are the energies of interacting monomers: the ordinary and exotic helium atom, respectively (cf. Fig.~\ref{fig:geometry}). 
All energies are computed in the so-called dimer-centered basis set~\cite{DCBS95}. 
This is equivalent to applying the counterpoise scheme~\cite{Boys1970} to remove the basis set superposition error (BSSE), which is a consequence of unphysical lowering of the monomer energies due to the presence of basis functions at both sites in calculations for the dimer. 
Note that, within this approach, the energy $E_\mathrm{A}$ is no longer constant and depends slightly on all three Jacobi coordinates. 
Similarly, $E_\mathrm{BC}(r)$ becomes a slowly varying function of $R$ and $\theta$. 
In practice, each calculation of the interaction energy as defined in Eq.~\eqref{eq:defVint} requires three different calculations using the same basis set of the dimer.

The results obtained with finite-size basis sets are extrapolated to the complete basis set (CBS) limit. 
We assume that the basis set truncation error of the calculated interaction energies vanishes with the increasing value of the basis set cardinal number $X$ as $1/X^3$. \cite{Halkier1998,Helgaker2008} 
The CBS limit can then be obtained using the following two-point formula,
\begin{equation}
    V[X-1,X]=V[X]+(X-1)^3\frac{V[X]-V[X-1]}{X^3-(X-1)^3},
\end{equation}
where $V[X-1]$ and $V[X]$ are energies calculated using basis sets with cardinal numbers $X-1$ and $X$, respectively, and $V[X-1,X]$ is the extrapolated value (the explicit dependence of $V$ on $R$, $r$, and $\theta$ has been omitted here for clarity).
To each geometry of the system, we assign a theoretical uncertainty of the interaction energy resulting from the extrapolation procedure employed by us. 
This uncertainty is conservatively estimated as the magnitude of the difference between the extrapolated value and the value computed with the largest basiśśśs set used in the extrapolation,
\begin{equation}\label{eq:extrap-err}
\Delta V[X-1,X]=\Big|V[X-1,X]-V[X]\Big|.
\end{equation}

\begin{figure}[!t]
\includegraphics[width=0.95\textwidth]{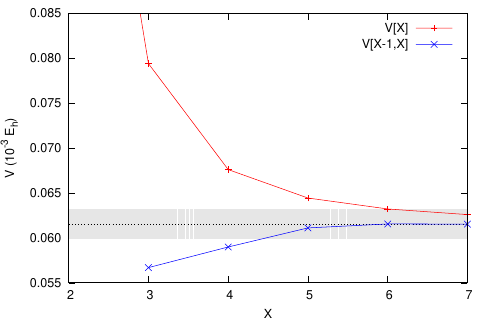}
\caption{Basis set convergence of the interaction energy for the $\apHef$--$\Hef$ system at the geometry $(R=5\,a_0, r=0.45\,a_0, \theta=60^\circ)$. Horizontal dotted line represents recommended value of the energy, and the shaded area is in the range of its estimated uncertainty.}
\label{fig:extrapolation}
\end{figure}

An example of the basis set convergence of the interaction energies, $V[X]$, calculated with the d$X$Z basis sets, as well as the convergence of the corresponding extrapolated values, $V[X-1,X]$, is illustrated in Fig.~\ref{fig:extrapolation}. 
The convergence of the unextrapolated results is smooth and monotonic. 
The values of the corresponding extrapolated ones seem to stabilize for $X\ge5$ and start to converge toward each other as the cardinal number $X$ increases.
Moreover, the $V[5,6]$ and $V[6,7]$ extrapolations are consistent with each other within their combined estimated error bars for 68 out of 70 grid points (97\% of cases) where the $V[7]$ results are available. 
These observations suggest that the extrapolation technique is able to provide reliable approximations to the CBS limit of the interaction energy for the $\apHef$--$\Hef$ system.
Additionally, we may safely assume the $V[5,6]$ extrapolants as our recommended values.
The mean fractional extrapolation error $\delta V_{\rm extr}=\langle \left|\Delta V[5,6]/V[5,6]\right| \rangle_{\rm grid}$, averaged over the whole set of 26505 grid points, is $0.96\times10^{-2}$. 

\subsection{Fitting of the potential}

Great care has been taken to provide a smooth approximation of the \emph{ab initio} points, that allowed us to overcome the problems that occurred with fitting of the previous $\apHef$--$\Hef$ PES. 
The resulting formula involves a sum of three terms, designed to fit the behavior of $V(R,r,\theta)$ at short, middle, and long interparticle distances, respectively, 
and smoothly damped outside the corresponding domain:
\begin{equation}\label{eq:Vlms}
V_{\rm fit}(R,r,\theta)=V_{\rm S}(R,r,\theta)+V_{\rm M}(R,r,\theta)+V_{\rm L}(R,r,\theta).
\end{equation}

The explicit form of the long-distance term $V_{\rm L}(R,r,\theta)$ is inferred from the analytical formula for the large-$R$ asymptotics of the interaction energy in a system consisting of a neutral $S$-state atom and an arbitrary $\Sigma$-state linear molecule.\cite{Buckingham_1967,Pack_1976} 
We include all the terms that vanish with $R$ as $R^{-10}$ or slower and the damping of $V_{\rm L}(R,r,\theta)$ in the short-$R$ regime is controlled with one adjustable non-linear parameter $q_1$, see Appendix 1 for details.

The short-distance term $V_{\rm S}(R,r,\theta)$ in Eq.~(\ref{eq:Vlms}) is taken in a form that accounts for the contribution of the Coulomb interaction in the pairs AB and AC (see Fig.~\ref{fig:geometry}) to the interaction energy, which dominates for small $R$:
\begin{equation}\label{eq:Vs}
V_{\rm S}(R,r,\theta)=\frac{4}{R_\mathrm{AB}}\left(1-f_1(q_2 R_\mathrm{AB})\right)-
\frac{2}{R_\mathrm{AC}}\left(1-f_1(q_2 R_\mathrm{AC})\right),
\end{equation}
where
\begin{equation}
R_\mathrm{AB}=\sqrt{R^2-2tRr\cos\theta+t^2r^2},\quad 
R_\mathrm{AC}=\sqrt{R^2+2(1-t)Rr\cos\theta+(1-t)^2r^2}.
\end{equation}
The damping of $V_{\rm S}(R,r,\theta)$ at large internuclear distances is achieved by means of the Tang--Toennies damping function\cite{Tang_1984} 
\begin{equation}\label{eq:TTdamp}
f_{\nu}(x) = 1 - e^{-x}\left(1+x+\frac{x^2}{2}+\cdots +\frac{x^\nu}{\nu!}\right)
\end{equation}
with $\nu=1$ and is controlled with the parameter $q_2$.

Finally, the mid-distance term $V_{\rm M}(R,r,\theta)$ in Eq.~(\ref{eq:Vlms}) is taken in the form of an expansion in a basis set $G=\{g(R,r,\theta),1540\}$ that is the tensor product of separate basis sets for each of the three Jacobi coordinates:
\begin{equation}
\{g(R,r,\theta),1540\}=\{g(R),14\}\otimes\{g(r),11\}\otimes\{g(\theta),10\}.
\label{eq:fullbasis}
\end{equation}
The integers in braces denote the dimension of the corresponding basis sets.
The $R$-basis is 
\begin{equation}
\{g(R), 14\}=\{g_k(R),k=1,\ldots,14\}=\{h(R,q_3,q_4)/ R^{k+1},k=1,\ldots , 14\},
\end{equation}
where the role of the damping factor $h(R,q_3,q_4)=1/\left(\exp(q_3(R-q_4))+1\right)$
is to suppress the contribution of $V_{\rm M}(R,r,\theta)$ at large $R$.
The $r$-basis is
\begin{equation}
\{g(r), 11\}=\{g_m(r),m=1,\ldots,11\}=\{ r^{m-1},\;m=1,\ldots ,11\},
\end{equation}
and the $\theta$-basis $\{g(\theta),10\}=\{g_l(\theta),l=1,\ldots,10\}$ consists of
\begin{align}
g_l(\theta)&=P_{l-1}(\cos\theta)),\ l=1,\ldots , 8,\\ 
g_9(\theta)&=\frac{1}{(\theta-q_5)^2+q_6}+\frac{1}{(\theta-360+q_5)^2+q_6}, \\
g_{10}(\theta)&=\frac{1}{(\theta-180)^2+q_6},
\end{align}
where $q_5$ and $q_6$ are adjustable parameters.
The explicit form of the  $r$- and $\theta$-bases provides compatibility with $V_{\rm L}(R,r,\theta)$.
The expansion of $V_{\rm M}(R,r,\theta)$ in $\{g(R,r,\theta),1540\}$ reads:
\begin{equation}\label{eq:Vm}
V_{\rm M}(R,r,\theta)=
\sum_{j\in G^*}p_{j}\,g_j(R,r,\theta)=
\sum_{j\in G^*}p_{j}\,g_{k_j}(R)\,g_{m_j}(r)\,g_{l_j}(\theta),
\end{equation}
where $G^*$ is a subset, described below, of the basis set in Eq.~(\ref{eq:fullbasis}).
Therefore, $V_{\rm M}(R,r,\theta)$ involves the set of linear parameters $p_j$ and 4 non-linear parameters $q_3,\ldots,q_6$. 

The subset $G^*$ in Eq.~(\ref{eq:Vm}) includes 1040 out of the 1540 elements of $G$ and is obtained by means of iterative filtration (equivalent to setting a part of the coefficients $p_j$  equal to zero).
The filtration process begins with setting $G^*=G$.
At each step, we optimize all 6 non-linear parameters appearing in $V_\mathrm{fit}(R,r,\theta)$ of Eq.~\eqref{eq:Vlms} using a procedure based on the simplex (Nelder--Mead) method \cite{Press_2007} and using $\chi^2=\sum_{i} (V_\mathrm{fit}(R_i, r_i,\theta_i)/V(R_i, r_i,\theta_i)-1)^2$ as the objective function, where the summation index $i$ runs over all grid points. 
Within the method, a linear determination of the parameters $p_j$ in Eq.~\eqref{eq:Vm} is performed in every simplex vertex.
Next, to quantify the importance of each basis function present in the set $G^*$, we calculate their norms, $||g_j||$, defined as $||g_j||=\sqrt{\sum_i g_j(R_i,r_i,\theta_i)^2}$.
Note that the norms depend on actual values of the non-linear parameters and have to be recalculated in every iteration. 
Finally, we find the minimal product $|p_j|\times||g_j||$ and exclude the corresponding $g_j(R,r,\theta)$ function from $G^*$. 
The procedure of filtration is stopped when further reduction of $G^*$ would significantly worsen the fit.
The main criterion for this is an abrupt increase of the number of grid points with more than $1\% $ relative error (in our case---more than 20 such points) which persist in next steps.
The filtration rules out, among others, all basis functions behaving as $\sim 1/R^2$ or $\sim 1/R^3$, and all but one basis functions behaving as $\sim 1/R^4$.
The values of the non-vanishing parameters of the final fit are given in Supplementary Materials~\cite{Supplementary_Materials}.

The smooth fit of Eqs.~(\ref{eq:Vlms}-\ref{eq:Vm}) provides fractional accuracy of the approximation that is fully consistent with the precision of the quantum chemistry calculations of the interaction energy.  
The mean fractional error of the fit $\delta V_{\rm fit}=\langle \left|V_{\rm fit}(R_i,r_i,\theta_i)/V(R_i,r_i,\theta_i)-1\right|\rangle_{\rm grid}$, averaged over the whole set of 26505 grid points, is $0.49\times10^{-3}$. 

\begin{figure}[!t]
    \centering
    \includegraphics[width=0.95\linewidth]{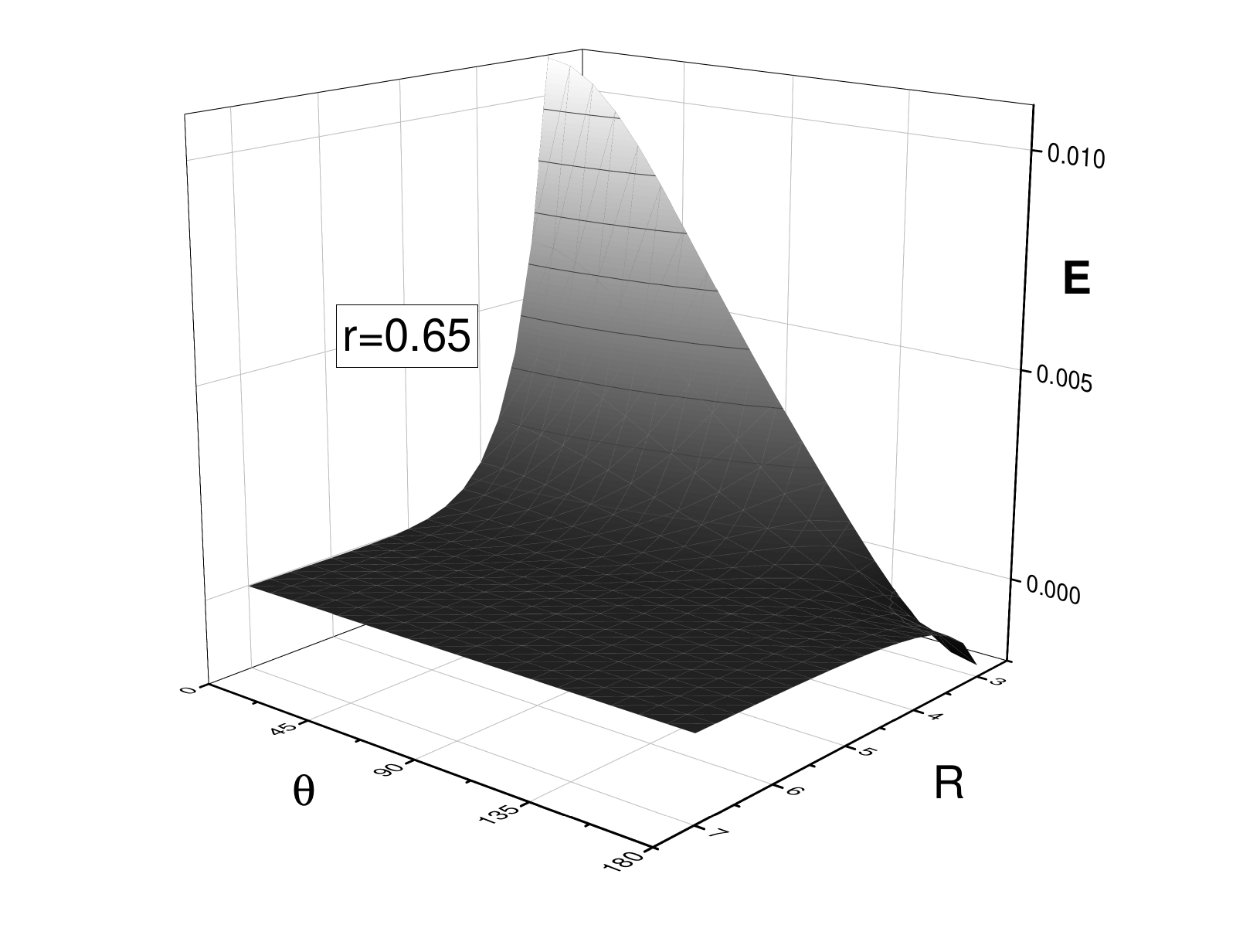}
    \caption{$\apHef$--$\Hef$ potential energy surface $V(R, r, \theta)$ for $r=0.65\,a_{0}$. The energy and distances are in atomic units, and the angles are in degrees.}
    \label{fig:pes-65}
\end{figure}

Fig.~\ref{fig:pes-65} presents an example 2D cut of the $\apHef$--$\Hef$ for $r=0.65\,a_{0}$, which is close to the average distance in the metastable states of $\apHef$. 
The presented surface differs from typical intermolecular potentials, which have a repulsive short-range wall. 
Here, if the antiproton (C in Fig.~\ref{fig:geometry}) faces the perturbing helium atom (A in Fig.~\ref{fig:geometry}), i.e., for $\theta = 180^{\circ}$, the electron density around the antiproton is insufficient to balance the attraction of the antiproton by the nucleus of the perturbing helium atom, leading to the presence of the attractive Coulomb-like well visible in Fig.~\ref{fig:pes-65}. 
This well enables the annihilation of the antiproton on the nucleus of the perturbing helium atom (A) rather than on the helium nucleus of $\apHef$ (B in Fig.~\ref{fig:geometry}). 
We note, however, that the effective state-dependent potentials, resulting from the average over $\theta$ and $r$, which enter quantum scattering calculations, do exhibit a repulsive wall at small $R$, as discussed later in Section~\ref{sec:results}.

We note that the analytical fit $V_{\mathrm{fit}}$ can be readily applied to antiprotonic helium-3 atoms, $\apHet$, as well as to other exotic-helium--helium systems involving pionic or kaonic helium.
In these cases, one simply replaces the mass ratio $t$ [defined in Eq.~\eqref{eq:t}] with the appropriate value $t'$ for the new system and performs a coordinate transformation that relates $(R',r',\theta')$ to the original $(R,r,\theta)$. 
This procedure is outlined in Appendix 2, ensuring that $V_{\mathrm{fit}}$ remains valid for any exotic helium atom once the coordinates are transformed properly.

\section{Fully quantum calculations of the collision-induced line-shape effects}

We calculate the pressure broadening and shift coefficients for helium-perturbed $\apHef$ spectral lines {using a fully quantum approach to describe the $\apHef$--$\Hef$ collisional process}. The exotic atoms, undergoing spectroscopic transitions between well-defined states, $a \rightarrow b$, are considered to be infinitely diluted in a thermal bath of helium atoms. Under the assumptions of the impact\footnote{The duration of each collision is significantly shorter than the time between successive collisions.} and binary\footnote{Each collision involves only a single exotic atom and a single perturbing helium atom.} approximations, the pressure broadening ($\gamma_{0}$) and pressure shift ($\delta_{0}$) coefficients can be derived from the scattering $S$-matrix elements resulting from two-body quantum scattering calculations. The two line-shape parameters are expressed in terms of the complex \emph{generalized spectroscopic cross-section}, $\sigma^{\kappa}(a,b;E_{\mathrm{kin}})$, as follows\cite{Wcislo_2021}
\begin{equation}
\label{eq:params} 
\gamma_{0} - i\delta_{0} = \frac{1}{2\pi c} \frac{\langle v_{r}\rangle}{k_{\mathrm{B}}T} \int_{0}^{+\infty} x e^{-x} \sigma^{\kappa}(a,b;E_{\mathrm{kin}}=xk_{\mathrm{B}}T) \,\mathrm{d}x ,
\end{equation}
where $c$ is the speed of light in vacuum, $k_{\mathrm{B}}$ is the Boltzmann constant, and  $T$ is the temperature. The term  $\langle v_{r}\rangle = \sqrt{8k_{\mathrm{B}}T/(\pi \mu)}$ represents the mean relative speed of the colliding pair, with $\mu$ being the reduced mass of the system. $\kappa$ denotes the tensorial rank of the radiation-matter interaction, which equals 1 for the electric dipole transitions considered in this work. 

Keeping the analogy between the exotic atom and a diatomic molecule, we now specify the meaning of the states $a$ and $b$. In the context of the exotic atom, these states are labeled by the hydrogen-like quantum numbers $n$ and $l$. To align with the diatomic molecule analogy, we relate these quantum numbers to the vibrational ($v$) and rotational ($j$) quantum numbers of a diatomic molecule via the following correspondence: $v=n-l-1$ and $j=l$. This allows us to treat the transition $a\rightarrow b$ as analogous to a \emph{rovibrational} transition in a diatomic molecule, $v_{a},j_{a} \rightarrow v_{b}, j_{b}$, and the corresponding formula for the generalized spectroscopic cross-section
\begin{equation}\label{eq:gsxs}
\begin{split}
 \sigma^{\kappa}(v_a,j_a, v_b,j_b; E_{\mathrm{kin}}) &= 
 \frac{\pi}{k^2} \sum_{J_a,J_b,L,L^{'}} (-1)^{L+L^{'}}\left[J_a, J_b\right]
\begin{Bmatrix}
 j_b & \kappa & j_{a}\\
 J_a & L & J_{b}
\end{Bmatrix}
\begin{Bmatrix}
 j_b & \kappa & j_{a}\\
 J_a & L' & J_{b}
\end{Bmatrix}
 \\ 
 &\times \Bigl(\delta_{LL^{'}} - S^{J_{a}}_{v_{a} j_{a} L^{'};v_{a} j_{a} L}(E_{\mathrm{kin}}+E_{v_{a}j_{a}}) \, S^{J_{b}\,*}_{v_{b} j_{b} {L}^{'};v_{b} j_{b} {L}}(E_{\mathrm{kin}}+E_{v_{b}j_{b}})\Bigr).
\end{split}
\end{equation}
Here, $L$ and $L'$ represent the pre- and post-collisional relative orbital angular momenta of the colliding pair (the end-over-end rotational angular momenta). $J_{a}$ and $J_{b}$ are the total angular momenta of the scattering system, arising from the coupling of the rotational angular momenta ($j_{a}$ and $j_{b}$, respectively) with $L$ or $L'$. $k =  \sqrt{2\mu E_{\mathrm{kin}}}$ is the wavevector (in atomic units), $[x_{1},x_{2}] = (2x_{1}+1)(2x_{2}+1)$, and terms in curly brackets denote 6-j symbols. Note that  the formula for the cross-section involves \emph{two} scattering $S$-matrices, both calculated at the same \emph{kinetic} energy, $E_{\mathrm{kin}}$, but corresponding to different \emph{total} energies: $E_{\mathrm{kin}}+E_{v_{a}j_{a}}$ and $E_{\mathrm{kin}}+E_{v_{b}j_{b}}$.

Scattering $S$-matrices are obtained by solving the quantum scattering problem for a system involving $\apHef$ and a helium atom. The Hamiltonian for the scattering system (in atomic units) is given by
\begin{equation}\label{eq:Hamiltonian}
{H} =  - \frac{1}{2\mu R} \frac{\partial^{2}}{\partial R^{2}} R + \frac{{\boldsymbol{L}}^{2}}{2\mu R^{2}} + {V}(R, r, \theta)  + {H}_{\mathrm{as}},
\end{equation}
where ${\boldsymbol{L}}^{2}$ represents the square of the end-over-end rotational angular momentum operator of the scattering system, ${V}(R, r, \theta)$ is the exotic-atom--helium-atom potential energy surface, and ${H}_{\mathrm{as}}$ is the asymptotic Hamiltonian describing colliding partners as $R\rightarrow \infty$. Since helium is treated as a structureless atom, ${H}_{\mathrm{as}}$ corresponds  to the Hamiltonian of the isolated $\apHef$, which is expressed as
\begin{equation}\label{eq:Hamiltonian_apHe}
{H}_{\mathrm{as}} = -\frac{1}{2\bar{m}r}\frac{\partial^{2}}{\partial r^{2}}r + \frac{{\boldsymbol{j}}^{2}}{{2\bar{m}r^{2}}} + {v}(r).
\end{equation}
Here $\bar{m}$ denotes the reduced mass of the exotic atom, ${\boldsymbol{j}}^{2}$ is the square of the rotational angular momentum operator, and ${v}(r)$ is the interaction energy of the exotic atom as a function of the distance between the antiproton and the helium-4 nucleus. The eigenvalues and eigenstates of the asymptotic Hamiltonian are denoted by $E_{vj}$ and $|v j m_j \rangle$, where $v$ is the vibrational quantum number, $j$ is the rotational quantum number, and $m_j$ is the projection of the rotational angular momentum on the space-fixed $Z$-axis. In the coordinate representation, the eigenstates are given by
\begin{equation}
\langle \vec{r} |vjm_{j}\rangle = \frac{\chi_{vj}(r)}{r}Y_{jm_{j}}(\hat{r})   ,
\end{equation}
where $Y_{jm_{j}}(\hat{r})$ 
is the spherical harmonic describing the orientation of the exotic atom in the space-fixed frame, and $\chi_{vj}(r)$ is the solution of the radial Schr\"{o}dinger equation for the isolated exotic atom
\begin{equation}
\label{eq:isolated_schrodinger}
   \Bigl( -\frac{d^{2}}{dr^{2}}+\frac{j(j+1)}{r^{2}} + 2 \bar{m}{v}(r) - 2\bar{m}E_{vj} \Bigr) \chi_{vj}(r) = 0 .
\end{equation}

The goal of the scattering calculations is to determine the eigenstates of the Hamiltonian in Eq.~\eqref{eq:Hamiltonian}, $|\Psi\rangle$, that correspond to a given total energy $E$, such that ${H}|\Psi\rangle = E|\Psi\rangle$. These eigenstates are expanded in a conveniently chosen basis set. A common choice is the space-fixed coupled basis, which combines the eigenvectors of the ${\boldsymbol{L}}^{2}$ operator, ${\boldsymbol{L}}^{2}|lm_{l}\rangle = l(l+1)|lm_{l}\rangle$, and the eigenvectors of the asymptotic Hamiltonian, $|vjm_j\rangle$,
\begin{equation}
\label{eq:SF_basis}
    |vjl JM \rangle = (-1)^{j+l-M} \sqrt{2J+1} \sum_{m_{j}, m_{l}}
    \begin{pmatrix}
    j & l & J \\    
    m_{j} & m_{l} & -M
    \end{pmatrix}
     |vjm_{j}\rangle |l m_{l}\rangle ,
\end{equation}
where the symbol in brackets is the 3-j symbol describing the coupling of the exotic atom's rotational angular momentum, ${\boldsymbol{j}}$, with the end-over-end angular momentum, ${\boldsymbol{L}}$, to form the total angular momentum, ${\boldsymbol{J}} = {\boldsymbol{j}} + {\boldsymbol{L}}$. The basis functions are also eigenvectors of the spatial parity operator, satisfying ${\Pi}|vjl JM\rangle = (-1)^{j+l}|vjl JM\rangle$.

Here, we adopt an alternative approach~\cite{Launay_1977}, where the basis set is defined as
\begin{equation}
    \label{eq:BF_basis}
    |vj\bar{\Omega}JM\epsilon \rangle = \frac{1}{\sqrt{2(1+\epsilon p \delta_{\bar{\Omega}, 0})}} \Bigl(|vj\Omega JM\rangle +{\epsilon p} |vj -\Omega JM\rangle\Bigr),
\end{equation}
with $|vj\Omega JM \rangle = |vj\Omega\rangle |JM\Omega\rangle$. In this representation,  $|vj\Omega\rangle$ are the eigenvectors of the exotic atom's rotational angular momentum operator, ${\boldsymbol{j}}^{2}|vj\Omega\rangle = j(j+1)|vj\Omega\rangle$, and its projection along the $R$-axis, ${j}_{R}|vj\Omega\rangle = \Omega |vj\Omega\rangle$. $|JM\Omega\rangle$ are the eigenvectors of the total angular momentum operator, ${\boldsymbol{J}}^{2}|JM\Omega\rangle = J(J+1)|JM\Omega\rangle$, its space-fixed $Z$-axis projection ${J}_{Z}|JM\Omega\rangle = M|JM\Omega\rangle$, and its projection on the $R$-axis ${J}_{R}|JM\Omega\rangle = \Omega |JM\Omega\rangle$. The basis functions in Eq.~\eqref{eq:BF_basis} are also eigenvectors of the spatial parity operator, satisfying ${\Pi}|vj\bar{\Omega}JM\epsilon\rangle = \epsilon |vj\bar{\Omega}JM \rangle$, where $\epsilon = \pm 1$. Additionally, $\bar{\Omega} = |\Omega|$, and $p = (-1)^{J}$.

Inserting the scattering wave function expanded in the basis introduced in Eq.~\eqref{eq:BF_basis}
\begin{equation}
\label{eq:wv_expansion}
    |\Psi \rangle = \sum_{v,j,\bar{\Omega},J,M,\epsilon}\frac{F_{vj\bar{\Omega}}^{JM{\epsilon}}(R)}{R} |vj\bar{\Omega}JM\epsilon\rangle ,
\end{equation}
into the Schr\"{o}dinger equation, and multiplying both sides by $\langle v'j'\bar{\Omega}'J'M'\epsilon' |$ leads to the set of coupled equations on the expansion coefficients, $F_{vj\bar{\Omega}}^{JM{\epsilon}}(R)$,
\begin{equation}\label{eq:CC_equations}
\frac{d^{2}}{dR^{2}}F_{vj\bar{\Omega}}^{JM\epsilon}(R) = \sum_{v'j'\bar{\Omega}'} W^{JM\epsilon}_{vj\bar{\Omega},v'j'\bar{\Omega}'} F_{v'j'\bar{\Omega}'}^{JM\epsilon}(R) .
\end{equation}
Since the total angular momentum, its projection on the space-fixed $Z$-axis, and parity all commute with the Hamiltonian in Eq.~\eqref{eq:Hamiltonian}, the equations are diagonal with respect to $J$, $M$ and $\epsilon$.  A key advantage of using the basis in Eq.~\eqref{eq:BF_basis} lies in the resulting structured coupling matrix, $W^{JM\epsilon}_{vj\bar{\Omega},v'j'\bar{\Omega}'} $\footnote{Here, $k_{vj} = \sqrt{2\mu (E-E_{vj})}$.}
\begin{equation}
\label{eq:coupling_matrix}
W^{JM\epsilon}_{vj\bar{\Omega},v'j'\bar{\Omega}'} = 2\mu \langle vj\bar{\Omega}JM\epsilon |{V}(R, r, \theta)|v'j'\bar{\Omega}'JM\epsilon\rangle + \frac{1}{R^{2}} \langle vj\bar{\Omega}JM\epsilon |{\boldsymbol{L}}^{2}|v'j'\bar{\Omega}'JM\epsilon\rangle - \delta_{vv'}\delta_{jj'}k_{vj}^{2} .
\end{equation}
The majority of the non-zero matrix elements of $W^{JM\epsilon}_{vj\bar{\Omega},v'j'\bar{\Omega}'}$ is grouped in blocks corresponding to fixed $\bar{\Omega}$ values. This is because the interaction potential couples only the states within the same $\bar{\Omega}$ block
\begin{equation}\label{eq:potential_term}
\begin{split}
\langle &vj\bar{\Omega}JM\epsilon |{V}(R, r, \theta)|v'j'\bar{\Omega}'JM\epsilon\rangle= \\
        &=  \delta_{\bar{\Omega}\bar{\Omega}'}(-1)^{\bar{\Omega}}\sqrt{[j,j']} \sum_{\lambda=|j-j'|}^{j+j'}A_{\lambda,vj,v'j'}(R) 
    \begin{pmatrix}
    j & j' & \lambda \\
    0 & 0 & 0 
    \end{pmatrix}
    \begin{pmatrix}
    j & j' & \lambda \\
    \bar{\Omega} & -\bar{\Omega} & 0 
    \end{pmatrix} .
\end{split}
\end{equation}
Here,  $A_{\lambda,vj,v'j'}(R)$ are the radial coupling terms of the potential energy surface
\begin{equation}\label{eq:radial_terms}
A_{\lambda,vj,v'j'}(R) = \frac{2\lambda+1}{2} \int_{0}^{\infty} \mathrm{d}r\, \chi_{vj}(r)\Bigl(\int_{0}^{\pi}\mathrm{d}\theta \sin{\theta}P_{\lambda}(\cos{\theta}){V}(R, r, \theta)\Bigr)\chi_{v'j'}(r) .
\end{equation}
The centrifugal term [the second term in Eq.~\eqref{eq:coupling_matrix}] introduces coupling between adjacent $\bar{\Omega}$ blocks and is diagonal in $j$. Matrix elements for this term can be found in Ref.~\cite{Launay_1977}.

Coupled equations are solved numerically using the renormalized Numerov algorithm~\cite{Johnson_1978} implemented in the \texttt{BIGOS} code developed in our group~\cite{Jozwiak_2024a, Jozwiak_2024_code}. At large $R$ values the scattering wave function is transformed to the space-fixed coupled basis defined in Eq.~\eqref{eq:SF_basis}. The scattering $S$-matrix is then determined by applying the appropriate boundary conditions to the scattering wave function at asymptotically large interatomic distances.

\section{\label{sec:details}Computational details}

\begin{figure}
    \centering
    \includegraphics[width=0.8\linewidth]{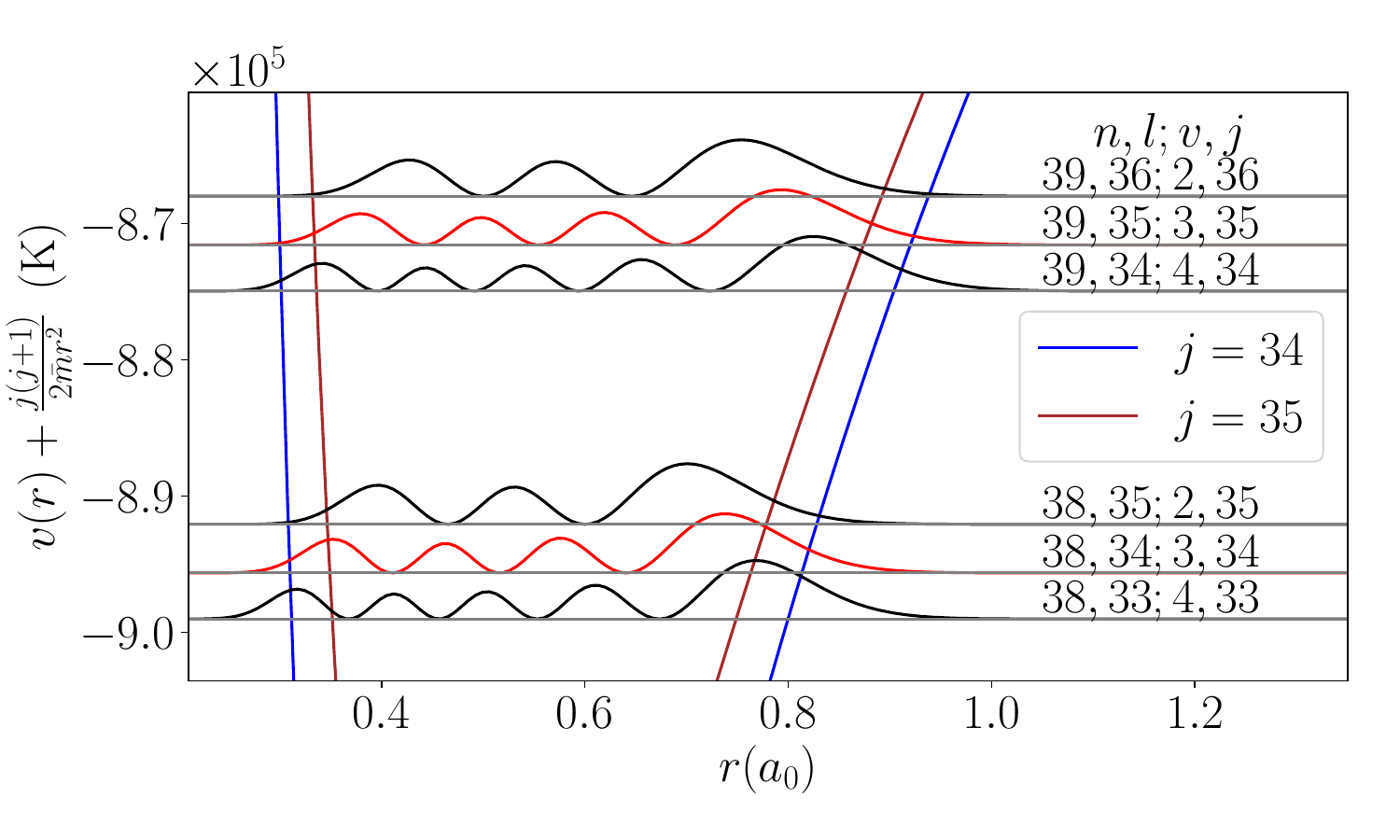}
    \includegraphics[width=0.8\linewidth]{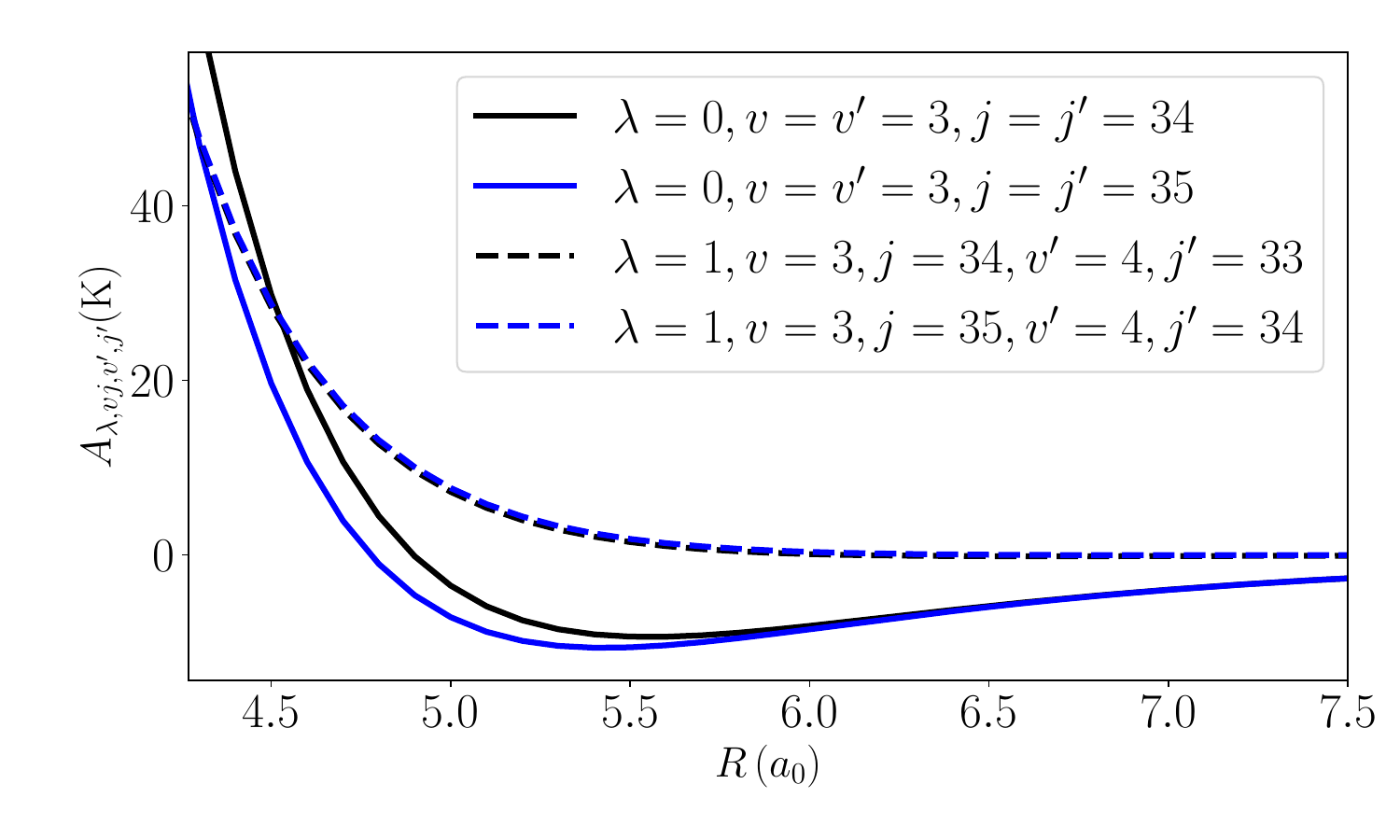}
    \caption{{{(Upper panel)}} Simplified energy diagram of $\apHef$, showing the energies and squared wave functions of the metastable states involved in the $v_{a}=3, j_{a}=35 \rightarrow v_{b}=3, j_{b}=34$ transition (highlighted in red), along with the two nearest neighboring states in energy for each {spectroscopic} level. 
    {{(Lower panel)}} State-dependent effective interaction potentials for the states of interest. Solid lines represent the isotropic terms ($\lambda = 0$), while dashed lines correspond to the strongest anisotropic term ($\lambda=1$) coupling states of interest to the nearest lower-lying state.}
    \label{fig:levels_coupling_terms}
\end{figure}

Quantum scattering calculations are performed for 29 states of antiprotonic helium-4, which allow us to calculate pressure broadening and shift coefficients for 50 transitions observed in this exotic atom. The full list of the transitions considered in this work is gathered in Table~\ref{tab:transitions}. Calculations are carried out for collision energies ranging from $10^{-3}$ to 150~K, which is sufficient to cover the entire energy distribution in the integral of Eq.~\eqref{eq:params}. To ensure convergence of the generalized spectroscopic cross-sections [Eq.~\eqref{eq:gsxs}] to within 0.1\%, the calculations include a sufficient number of $J$-blocks, ranging from 35 to 100 across the collision energy range. Radial expansion coefficients of the scattering wave function, $F_{vj\bar{\Omega}}^{JM{\epsilon}}(R)$, are propagated on an equidistant grid from $R_{\mathrm{min}} = 2.5\,a_{0}$ to $R_{\mathrm{max}} = 50\,a_{0}$ with 25 steps per half-de Broglie wavelength. 

Due to the significant energy spacing between eigenstates in $\apHef$ (of an order of $10^{3}$~K for the states considered), the number of rovibrational levels involved in the expansion of the scattering wave function [Eq.~\eqref{eq:wv_expansion}] is limited to the nearest neighboring states. Specifically, for scattering calculations involving antiprotonic helium in the $v, j$ state, only the states $v + 1, j-1 $, $v,j$, and $v -1, j +1$ were included in the calculations. This is illustrated for the case of the $v_{a}=3, j_{a}=35 \rightarrow v_{b}=3, j_{b}=34$ transition in the upper panel of Fig.~\ref{fig:levels_coupling_terms}. 

The radial coupling terms of the potential energy surface [Eq.~\eqref{eq:radial_terms}], which define the coupling matrix [Eq.~\eqref{eq:coupling_matrix}], are calculated numerically using Gauss-Legendre quadrature for the integral over $\theta$ and the Simpson rule for the integral over $r$. While transitions involving high $j$ values could, in principle, require summing over wide $\lambda$ range and could extend to very large values, these contribute negligibly to transitions between rotational levels differing by $\Delta j = \pm 1$. Consequently, the final calculations include coupling terms for $\lambda$ ranging from 0 to 7. The lower panel of Fig.~\ref{fig:levels_coupling_terms} presents the isotropic ($\lambda=0$) and the strongest anisotropic ($\lambda=1$) terms used in the calculations for the $v_{a}=3, j_{a}=35 \rightarrow v_{b}=3, j_{b}=34$ transition. The radial wave functions of the $\apHef$, $\chi_{vj}(r)$, used to average the potential energy surface over the exotic atom internal coordinate $r$, are obtained by solving Eq.~\eqref{eq:isolated_schrodinger} with the potential energy curve of Shimamura~\cite{Shimamura_1992} using the discrete variable representation-finite basis representation method.

\section{\label{sec:results}Results}

We determined the pressure broadening and shift coefficients for {50} helium-perturbed transitions in antiprotonic helium-4 (see Table~\ref{tab:transitions}) over a temperature range of 1.3~K to 10~K. Most measurements involving gaseous helium are conducted within the 4.5--10~K range; however, we extended our calculations to lower temperatures to account for experiments involving superfluid helium~\cite{Soter_2022}. Although the binary approximation is no longer valid in non-gaseous phases, the results presented here may still provide useful estimates for models of spectral broadening in the liquid, solid, and superfluid states of helium, serving as a baseline for the background contribution from binary collisions. The transitions are categorized into two types: those that do not involve a change in vibrational quantum number ($\Delta v=0$) and those that do ($\Delta v = 2$). In exotic atom spectroscopy, these two categories are referred to as favored ($\Delta n = \Delta l = \pm 1$) and unfavored ($\Delta n = -\Delta l = \pm 1$) transitions, respectively. The full tabulated values of pressure broadening and shift coefficients are provided in Supplementary Materials~\cite{Supplementary_Materials}.

\begin{table}[!t]
    -\centering
    \begin{tabular}{c|c|c|c}
        \multicolumn{2}{c|}{$\Delta v = 0$ transitions} &  \multicolumn{2}{c}{$\Delta v = 2$ transitions}  \\
        \hline
        $v_{a},j_{a} \rightarrow v_{b},j_{b}$ & Reference  & $v_{a},j_{a} \rightarrow v_{b},j_{b}$ & Reference \\ \hline
     $4,35 \rightarrow 4,34$ & \cite{Bakalov_2000, Hori_2006, Bakalov_2012, Hori_2016} &  $3,36 \rightarrow 5,35$ & \cite{Kobayashi_2013} \\
     $4,34 \rightarrow 4,33$ & \cite{Bakalov_2012} & $3,35 \rightarrow 5,34$ & \cite{Bakalov_2012} \\
     $3,36 \rightarrow 3,35$ & \cite{Bakalov_2012} & $3,34 \rightarrow 5,33$ & \cite{Bakalov_2012} \\
     $3,35 \rightarrow 3,34$ & \cite{Bakalov_2000, Hori_2001, Hori_2006, Bakalov_2012, Hori_2016} & $2,36 \rightarrow 4,35$ & \cite{Bakalov_2012}   \\
     $3,34 \rightarrow 3,33$ & \cite{Bakalov_2012} & $2,35 \rightarrow 4,34$ & \cite{Hori_2016}     \\
     $3,33 \rightarrow 3,32$ & \cite{Bakalov_2012} & $2,34 \rightarrow 4,33$ & \cite{Bakalov_2000, Hori_2001}  \\
     $2,37 \rightarrow 2,36$ & \cite{Bakalov_2012} & $2,33 \rightarrow 4,32$ & \cite{Bakalov_2012} \\
     $2,36 \rightarrow 2,35$ & \cite{Bakalov_2000, Bakalov_2012} &  $1,37 \rightarrow 3,36$ & \cite{Bakalov_2012}    \\
     $2,35 \rightarrow 2,34$ & \cite{Bakalov_2000, Bakalov_2012} &  $1,36 \rightarrow 3,35$ & \cite{Bakalov_2012} \\
     $2,34 \rightarrow 2,33$ & \cite{Bakalov_2000, Hori_2006, Bakalov_2012, Hori_2016} &$1,35 \rightarrow 3,34$ & \cite{Bakalov_2000, Hori_2001,  Hori_2006, Bakalov_2012, Hori_2016} \\
     $2,33 \rightarrow 2,32$ & \cite{Bakalov_2012} & $1,34 \rightarrow 3,33$ & \cite{Hori_2001, Bakalov_2012}   \\
     $2,32 \rightarrow 2,31$ & \cite{Bakalov_2012} & $1,33 \rightarrow 3,32$ & \cite{Bakalov_2012}  \\
     $1,38 \rightarrow 1,37$ & \cite{Bakalov_2012} & $1,32 \rightarrow 3,31$ & \cite{Bakalov_2012} \\
     $1,37 \rightarrow 1,36$ & \cite{Bakalov_2000, Bakalov_2012} & $0,38 \rightarrow 2,37$ & \cite{Bakalov_2012} \\
     $1,36 \rightarrow 1,35$ & \cite{Bakalov_2000, Bakalov_2012} & $0,37 \rightarrow 2,36$ &  \cite{Bakalov_2012}  \\
     $1,35 \rightarrow 1,34$ & \cite{Bakalov_2012} & $0,36 \rightarrow 2,35$ & \cite{Bakalov_2012}    \\
     $1,34 \rightarrow 1,33$ & \cite{Bakalov_2000, Hori_2006, Bakalov_2012}  & $0,35 \rightarrow 2,34$ & \cite{Bakalov_2012}\\
     $1,33 \rightarrow 1,32$ & \cite{Bakalov_2000, Hori_2001, Hori_2006, Bakalov_2012, Hori_2016} & $0,34 \rightarrow 2,33$ & \cite{Bakalov_2012}  \\
     $1,32 \rightarrow 1,31$ & \cite{Bakalov_2012} &  $0,33 \rightarrow 2,32$ & \cite{Bakalov_2012, Hori_2016}  \\
     $1,31 \rightarrow 1,30$ & \cite{Bakalov_2012} &  $0,32 \rightarrow 2,31$ & \cite{Bakalov_2012} \\
     $0,39 \rightarrow 0,38$ & \cite{Bakalov_2012} &  $0,31 \rightarrow 2,30$ & \cite{Bakalov_2012}  \\
     $0,38 \rightarrow 0,37$ & \cite{Bakalov_2000, Bakalov_2012} &    &   \\
     $0,37 \rightarrow 0,36$ & \cite{Bakalov_2000, Bakalov_2012} &    &   \\
     $0,36 \rightarrow 0,35$ & \cite{Bakalov_2012} &    &  \\
     $0,35 \rightarrow 0,34$ & \cite{Bakalov_2012} &    &  \\
     $0,34 \rightarrow 0,33$ & \cite{Bakalov_2012} &    &  \\
     $0,33 \rightarrow 0,32$ & \cite{Bakalov_2012} &    &  \\
     $0,32 \rightarrow 0,31$ & \cite{Hori_2001, Bakalov_2012} &    &   \\
     $0,31 \rightarrow 0,30$ & \cite{Bakalov_2000, Hori_2006, Bakalov_2012, Hori_2016} &    &   \\
     \hline
    \end{tabular}
    \caption{List of transitions considered in this work, along with references to previous experimental and theoretical studies.}
    \label{tab:transitions}
\end{table} 

\begin{figure}[!t]
    \centering
    \includegraphics[width=\linewidth]{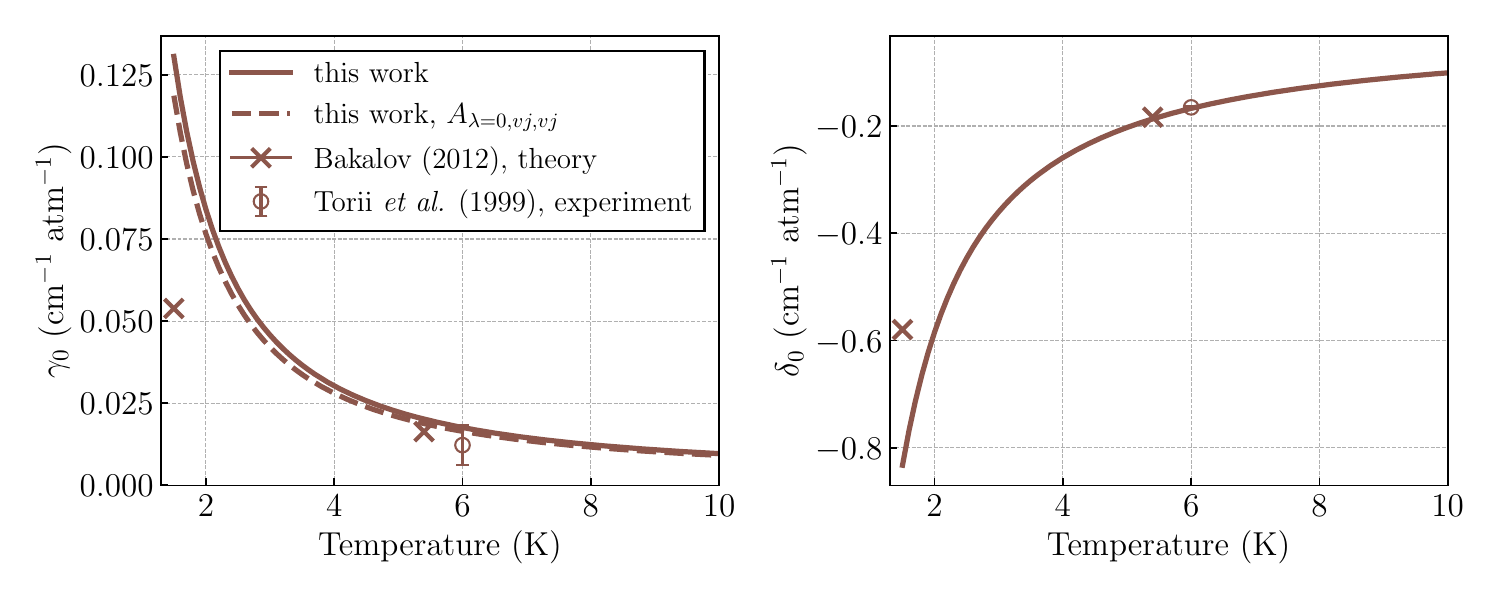}
    \caption{Pressure broadening ($\gamma_{0}$) and pressure shift ($\delta_{0}$) coefficients for the $v_{a}=3, j_{a}=35 \rightarrow v_{b}=3, j_{b}=34$ transition in $\apHef$. Dashed lines correspond to the results of the calculations performed only on the isotropic part of the interaction potential [$A_{\lambda=0,vj,vj}$ in Eq.~\eqref{eq:potential_term}].}
    \label{fig:representative_transition}
\end{figure}

Given the large number of transitions studied, we begin by focusing on a representative case for which both experimental and theoretical reference data are available: the favored $v_{a}=3, j_{a}=35 \rightarrow v_{b}=3, j_{b}=34$ transition, see the upper panel in Fig.~\ref{fig:levels_coupling_terms}. Pressure broadening and shift coefficients for this line are presented in Fig.~\ref{fig:representative_transition}. We find excellent agreement with the experimental values of both line-shape parameters reported in Ref.~\cite{Torii_1999}, and a surprisingly good agreement with the semiclassical calculations at $T=5.4$~K~\cite{Bakalov_2000,Bakalov_2012}. This is particularly interesting, given that the reference theoretical calculations employed a $\apHef$--$\Hef$ PES computed on a limited grid of \emph{ab initio} points, relied on Anderson's semiclassical line-shape formalism, and essentially neglected anisotropic effects of the interaction. However, the semiclassical results predict a less steep temperature dependence of both parameters, leading to a factor-of-two underestimation of $\gamma_{0}$ and 40\% underestimation of $\delta_{0}$ at $T=1.5$~K.

The representative transition also illustrates the key mechanisms that determine pressure broadening and shift in helium-perturbed $\apHef$ lines. First, we note that, contrary to most molecular systems, the magnitude of the pressure broadening coefficients is much smaller than that of the pressure shift coefficients. To explore this behavior in more detail, it is essential to recognize that the pressure broadening coefficient involves two contributions: the inelastic and dephasing terms~\cite{BenReuven_1966}. The inelastic contribution arises from the half-sum of the inelastic rate coefficients from the two spectroscopic states involved (see Eq.~A.10 in Ref.~\cite{Thibault_2016}). However, in antiprotonic helium, the large separation between energy levels makes the inelastic contribution negligible. For collision energies considered here, the inelastic quenching from $v, j$ to the $v + 1, j-1 $ state makes only $10^{-6}$ of $\gamma_{0}$. Thus, pressure broadening in $\apHef$ is almost entirely determined by the dephasing contribution, which stems from the differences in the elastic scattering amplitudes in the two levels involved in the spectroscopic transition.

In typical molecular systems, the dephasing contribution is significant for rovibrational transitions, because the effective interaction potentials for states $v_{a}$ and $v_{b}$ can differ substantially. For the chosen representative transition, as well as for all favored transitions considered in this work, the effective interaction potentials belong to the \emph{same} vibrational manifold but differ in rotational quantum number. Therefore, the pressure broadening of the helium-perturbed $\apHef$ transitions provides an example of broadening by \emph{rotational} dephasing. This dephasing results from reorienting collisions i.e., changes in the orientation of the rotational angular momentum $\boldsymbol{j}$) induced by the anisotropic components of the interaction potential [$\lambda \neq 0$ in Eq.~\eqref{eq:potential_term}], and purely dephasing collisions driven by the isotropic components of the interaction potential.

To estimate the relative importance of these factors, following Ref.~\cite{Thibault_2016}, we performed quantum scattering calculations for the representative $v_{a}=3, j_{a}=35 \rightarrow v_{b}=3, j_{b}=34$ transition, neglecting all anisotropic components of the PES. In other words, only the $\lambda=0$ term was kept in the expansion of the interaction potential in Eq.~\eqref{eq:potential_term}. The resulting broadening coefficient, shown as the dashed line in the left panel of Fig.~\ref{fig:representative_transition}, was 5--8\% lower than the original value across the temperature range of 1.3--10~K. This indicates that more than 90\% of the broadening is driven by purely phase-changing collisions arising from differences in the isotropic interaction potentials, $A_{\lambda=0,v_{a}j_{a},v_{a}j_{a}}$ and $A_{\lambda=0,v_{b}j_{b},v_{b}j_{b}}$ (see the lower panel of Fig.~\ref{fig:levels_coupling_terms}). The minimal role of anisotropy in this case also helps explain the success of semiclassical calculations of the pressure broadening coefficients~\cite{Bakalov_2000, Bakalov_2012}, which rely only on the isotropic parts of the potential energy surface.

\subsection{$\Delta v=0$ transitions}

\begin{figure}[!t]
    \centering
    \includegraphics[width=\linewidth]{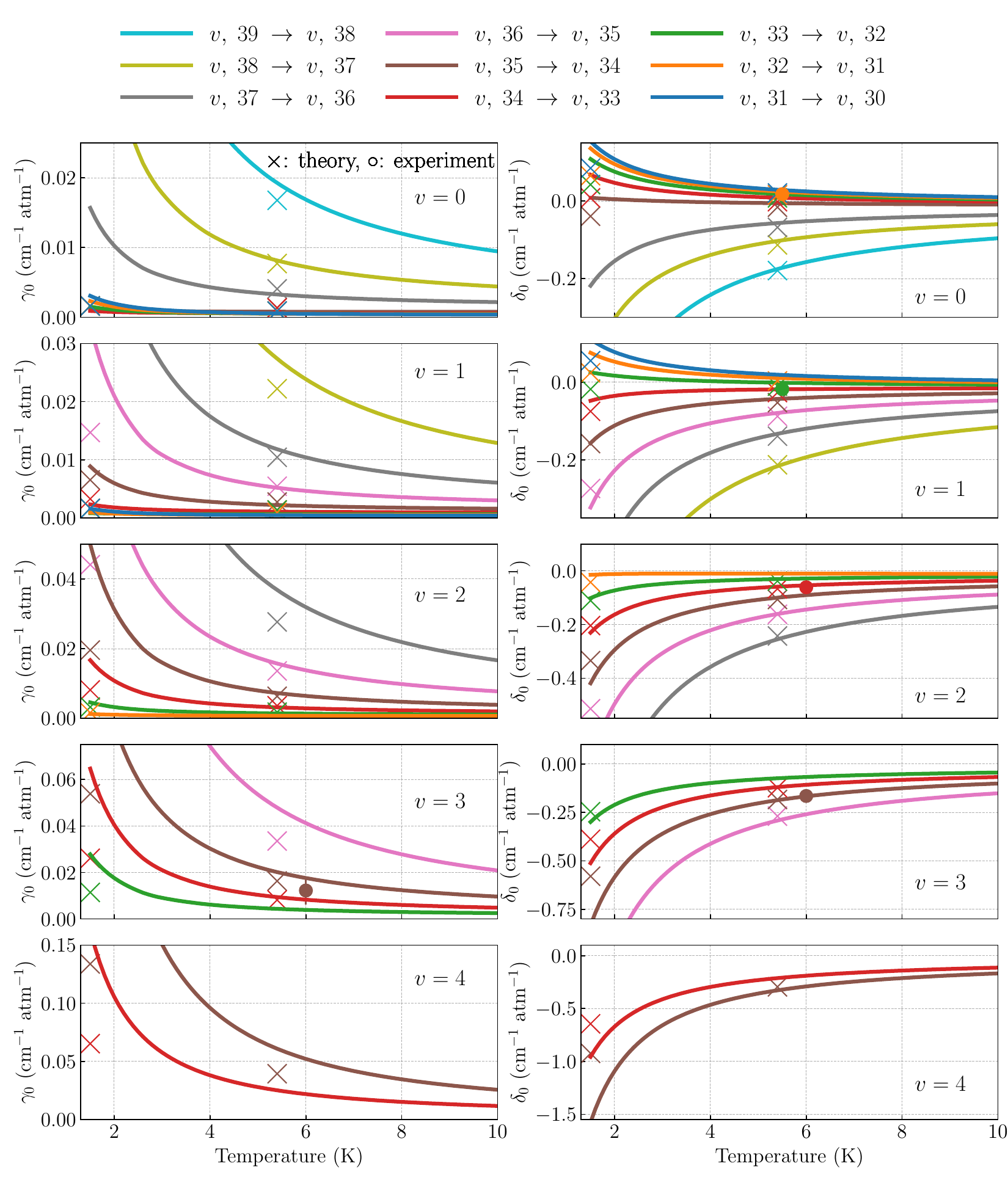}
    \caption{Pressure broadening ($\gamma_{0}$) and pressure shift ($\delta_{0}$) coefficients for favored ($\Delta v =0$) transitions in $\apHef$.}
    \label{fig:favoured_transitions}
\end{figure}

We present the complete set of results for favored transitions ($\Delta v=0$) in Fig.~\ref{fig:favoured_transitions}. Each row corresponds to a different vibrational manifold ($v=0$ through $v=4$), and each panel shows the computed temperature dependence of $\gamma_0$ (left) and $\delta_0$ (right).

We find overall satisfactory agreement with the experimental pressure shift coefficients. As noted earlier, the only experimental data for $\gamma_{0}$ was discussed for the representative transition. In comparison with previous calculations, we observe a good agreement for both $\gamma_{0}$ and $\delta_{0}$ at $T=5.4$~K, particularly for the highest rotational transitions in each vibrational manifold. Similar to the $v_{a}=3, j_{a}=35 \rightarrow v_{b}=3, j_{b}=34$ line discussed earlier, our results reveal a more pronounced temperature dependence of both line-shape parameters than that predicted by semiclassical calculations. This results in a systematic underestimation of $\gamma_{0}$, and, in some cases, even an opposite sign of $\delta_{0}$ at $T=1.5$~K.

Our results also reveal a clear, and more pronounced than in semiclassical calculations, trend over the 1.5--10~K range: the pressure broadening coefficient increases with $j$. This may seem counterintuitive, as in typical molecular systems $\gamma_{0}$ tends to \emph{decrease} with $j$ for large rotational quantum numbers, due to the widening of rotational level spacings, which suppresses inelastic transitions. Meanwhile, the phase-changing contribution remains relatively constant since the isotropic interaction potentials change only slightly with $j$.
In antiprotonic helium, however, the regime is fundamentally different. Inelastic collisions are absent for all $j$, and the isotropic interaction potentials vary significantly with rotational quantum number. These large differences in isotropic potentials also explain the magnitude of the pressure shift coefficients.

Interestingly, within a single vibrational manifold, some transitions exhibit pressure shifts of opposite signs. While we cannot fully explain this behavior, our extended calculations at $T > 10$~K indicate that all $\Delta v =0$ transitions ultimately have negative pressure shifts, with magnitudes increasing with $j$.

\subsection{$\Delta v=2$ transitions}

\begin{figure}[!t]
    \centering
    \includegraphics[width=\linewidth]{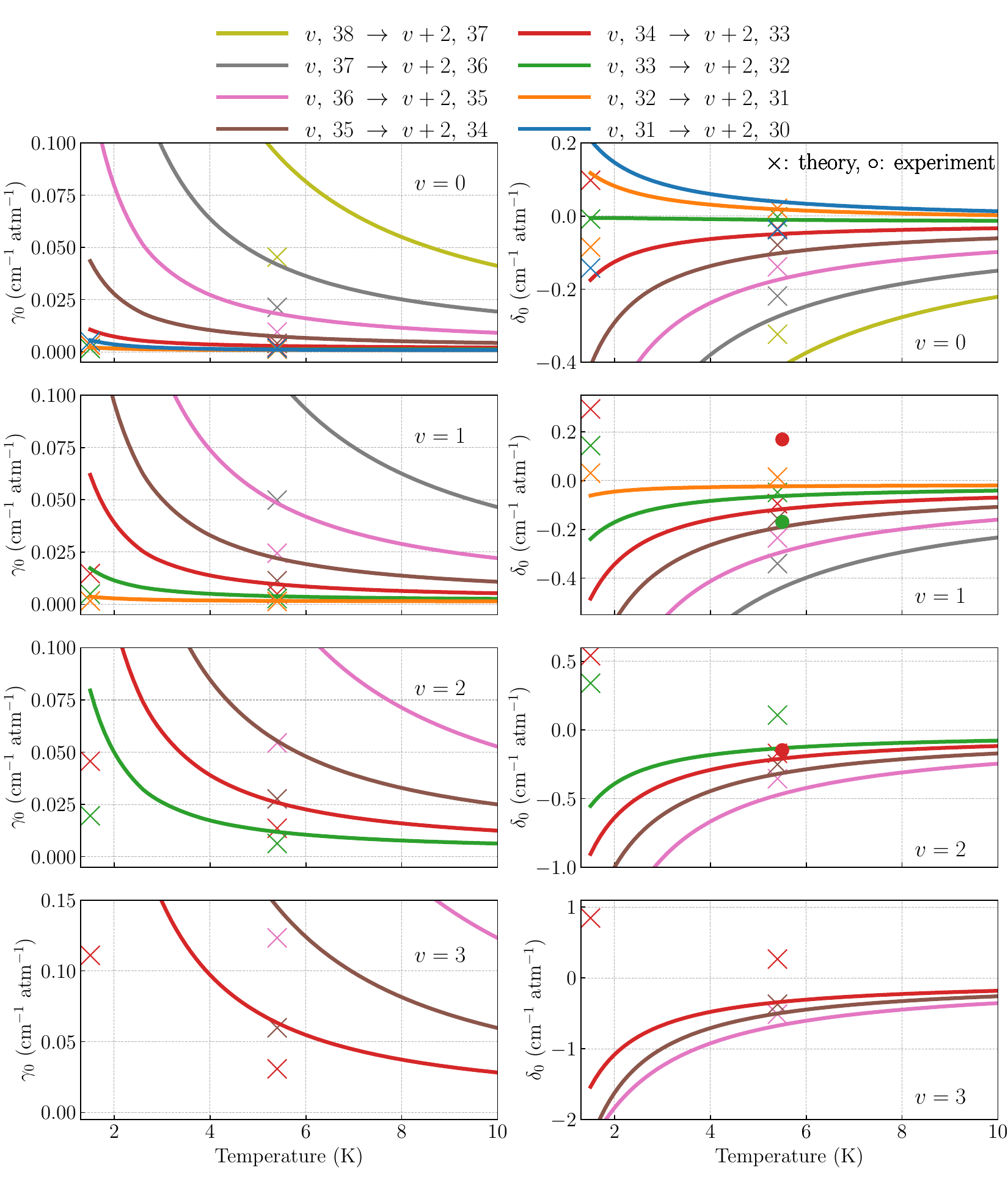}
    \caption{Pressure broadening ($\gamma_{0}$) and pressure shift ($\delta_{0}$) coefficients for unfavored ($\Delta v = 2$) transitions in $\apHef$.}
    \label{fig:unfavoured_transitions}
\end{figure}

Fig.~\ref{fig:unfavoured_transitions} presents the pressure broadening (left column) and pressure shift (right column) coefficients for helium-perturbed $\Delta v=2$ transitions in $\apHef$ considered in this work. Transitions originating from levels with $v_{a} =0$, 1, 2, 3 are shown in separate rows of the figure.

Compared to favored transitions, we observed more pronounced discrepancies with both the semiclassical predictions and the experimental pressure shifts. In particular, the $v_{a}=1, j_{a}=34 \rightarrow v_{b}=3, j_{b}=33$ transition exhibits a surprising disagreement with the experimental shift reported in Ref.~\cite{Hori_2001}---a discrepancy that had already been noted relative to semiclassical calculations in Ref.~\cite{Bakalov_2000}. This inconsistency remains unexplained in the framework of our quantum scattering calculations performed on the improved $\apHef$--$\Hef$ potential energy surface. As in the $\Delta v=0$ case, our results show a stronger dependence of both $\gamma_{0}$ and $\delta_{0}$ on temperature and rotational quantum number than semiclassical calculations, leading to significant differences at $T=1.5$~K.

In terms of general trends, we again find that the magnitude of the pressure shift coefficients exceeds that of the pressure broadening coefficients. {Across all vibrational manifolds} the unfavored transitions exhibit significantly larger broadening than their favored counterparts originating from the same initial state, $v_{a}, j_{a}$. This enhancement can be attributed to \emph{vibrational} dephasing, which is absent in $\Delta v = 0$ transitions. Vibrational dephasing arises from differences in the effective isotropic interaction potentials between \emph{different} vibrational manifolds.

Similar to $\Delta v = 0$ transitions, some $\Delta v = 2$ transitions exhibit positive pressure shifts at $T<10$~K. While we cannot definitively explain this behavior for these specific lines, we note that at higher temperatures (not shown in the figure), all transitions considered exhibit negative pressure shifts, with magnitudes increasing with $j$.

\section{Conclusion}

We have performed the first fully \emph{ab initio} calculations of the collisional perturbation of spectral lines in antiprotonic helium. To overcome the limitations of previous studies, we developed a new, state-of-the-art $\apHef$--$\Hef$ potential energy surface, allowing us to address the range of intermolecular geometries that are sufficient to describe all experimentally relevant transitions in exotic helium atoms.

Using the new PES, we performed rigorous quantum scattering calculations for the $\apHef$--$\Hef$ system and computed scattering $S$-matrices to determine pressure broadening and shift coefficients for 50 electric dipole transitions in $\apHef$ over a wide range of temperatures. Our results are in good agreement with the limited available experimental data and provide the first rigorous benchmark of previous semiclassical line-shape calculations.

This dataset provides a valuable reference for high-precision spectroscopic measurements in $\apHef$, which are used to test three-body QED calculations and test the fundamental CPT symmetry. The new PES and methodology presented here also pave the way for similar studies in other exotic atom systems, including the $\apHet$ isotope of antiprotonic helium, as well as pionic and kaonic helium atoms. These calculations could support future precision measurements aimed at determining the masses of exotic particles with improved accuracy.

Our study covers temperatures as low as 1.5~K. While the binary collision approximation breaks down in non-gaseous phases, the results reported here can still serve as fully quantum estimates (as opposed to earlier, semiclassical calculations)\cite{Adamczak_2013, Adamczak_2014} for spectral broadening in liquid, solid, and superfluid helium. This is particularly relevant for interpreting precision measurements of antiprotonic helium transitions performed in non-gaseous environments.

\section{Acknowledgments}

This paper was initiated by Prof. Bogumił Jeziorski, who actively participated in all stages of the work until his premature passing. His interest in the field dates back to the 1990s, when he applied jointly with Prof. Krzysztof Szalewicz the symmetry-adapted perturbation theory formalism to the calculation of the first potential energy surface for the interaction of ordinary with exotic helium atoms. He anticipated the perspectives of extending the investigations to a broader range of phenomena with exotic atoms, and recent developments in the field have confirmed the correctness of these predictions.

The research is funded by the European Union (ERC-2022-STG, H2TRAP, 101075678). Views and opinions expressed are however those of the author(s) only and do not necessarily reflect those of the European Union or the European Research Council Executive Agency. H. J. was supported by the National Science Centre in Poland through Project No. 2024/53/N/ST2/02090. D. B. and M. S. are supported by BNSF grant KP-06-N58/5. We gratefully acknowledge the Polish high-performance computing infrastructure PLGrid (HPC Centers: ACK Cyfronet AGH) for providing computer facilities and support within the computational grant, Grant No. PLG/2025/018511. Created out using resources provided by Wroclaw Centre for Networking and Supercomputing (http://wcss.pl).

\newpage
\section{Appendices}

\subsection{The explicit form of the long-distance term in the PES fit}

The large-$R$ asymptotic behavior of the interaction energy for a system depicted in Fig.~\ref{fig:geometry} has the following form:\cite{Buckingham_1967,Pack_1976} 
\begin{equation}\label{VLR}
\begin{split}
V(R,\theta,r)
&\sim\left(B_{6,0}(r)+B_{6,2}(r)P_2(\cos\theta)\right)R^{-6} \\
&+\left(B_{7,1}(r)P_1(\cos\theta)+B_{7,3}(r)P_3(\cos\theta)\right)R^{-7} \\
&+\left(B_{8,0}(r)+B_{8,2}(r)P_2(\cos\theta)+B_{8,4}(r)P_4(\cos\theta)\right)R^{-8} \\
&+\left(B_{9,1}(r)P_1(\cos\theta)+B_{9,3}(r)P_3(\cos\theta)+B_{9,5}(r)P_5(\cos\theta)\right)R^{-9} \\
&+O(R^{-10}),
\end{split}
\end{equation}
where $P_l(\cos\theta)$ are the Legendre polynomials. The $r$-dependent coefficients $B_{n,l}(r)$ are defined in terms of static $2^k$-pole polarizabilities of A, $\alpha^\mathrm{A}_k$, permanent $2^k$-pole moments of BC, $Q^\mathrm{BC}_k(r)$, and dispersion coefficients $C_{n,l}(r)$:
\begin{equation}
  \label{eq:Bs}
  \begin{split}
    B_{6,0}(r)&= -\left(\alpha_1^\mathrm{A}\,Q_1^\mathrm{BC}(r)^2+C_{6,0}(r)\right), \\
    B_{6,2}(r)&= -\left(\alpha_1^\mathrm{A}\,Q_1^\mathrm{BC}(r)^2+C_{6,2}(r)\right), \\[2ex]
    B_{7,1}(r)&= -\left(-\frac{18}{5}\,\alpha_1^\mathrm{A}\,Q_1^\mathrm{BC}(r)Q_2^\mathrm{BC}(r)+C_{7,1}(r)\right), \\
    B_{7,3}(r)&= -\left(-\frac{12}{5}\,\alpha_1^\mathrm{A}\,Q_1^\mathrm{BC}(r)Q_2^\mathrm{BC}(r)+C_{7,3}(r)\right), \\[2ex]
    B_{8,0}(r)&= -\left(\frac{3}{2}\,\alpha_1^\mathrm{A}\,Q_2^\mathrm{BC}(r)^2+\frac{5}{2}\,\alpha_2^\mathrm{A}\,Q_1^\mathrm{BC}(r)^2+C_{8,0}(r)\right), \\
    B_{8,2}(r)&= -\left(\frac{36}{7}\,\alpha_1^\mathrm{A}\,Q_1^\mathrm{BC}(r)Q_3^\mathrm{BC}(r)+\frac{12}{7}\,\alpha_1^\mathrm{A}\,Q_2^\mathrm{BC}(r)^2+2\,\alpha_2^\mathrm{A}\,Q_1^\mathrm{BC}(r)^2+C_{8,2}(r)\right), \\
    B_{8,4}(r)&= -\left(\frac{20}{7}\,\alpha_1^\mathrm{A}\,Q_1^\mathrm{BC}(r)Q_3^\mathrm{BC}(r)+\frac{9}{7}\,\alpha_1^\mathrm{A}\,Q_2^\mathrm{BC}(r)^2+C_{8,4}(r)\right), \\[2ex]
    B_{9,1}(r)&= -\left(-\frac{36}{7}\,\alpha_1^\mathrm{A}\,Q_2^\mathrm{BC}(r)Q_3^\mathrm{BC}(r)-12\,\alpha_2^\mathrm{A}\,Q_1^\mathrm{BC}(r)Q_2^\mathrm{BC}(r)+C_{9,1}(r)\right), \\
    B_{9,3}(r)&= -\left(-\frac{20}{3}\,\alpha_1^\mathrm{A}\,Q_1^\mathrm{BC}(r)Q_4^\mathrm{BC}(r)-4\,\alpha_1^\mathrm{A}\,Q_2^\mathrm{BC}(r)Q_3^\mathrm{BC}(r)-6\,\alpha_2^\mathrm{A}\,Q_1^\mathrm{BC}(r)Q_2^\mathrm{BC}(r)+C_{9,3}(r)\right),  \\
    B_{9,5}(r)&= -\left(-\frac{10}{3}\,\alpha_1^\mathrm{A}\,Q_1^\mathrm{BC}(r)Q_4^\mathrm{BC}(r)-\frac{20}{7}\,\alpha_1^\mathrm{A}\,Q_2^\mathrm{BC}(r)Q_3^\mathrm{BC}(r)+C_{9,5}(r)\right).
  \end{split}
\end{equation}
The values of $B_{n,l}(r)$ were calculated at $D=360$ distances $r$ ranging from 0.01 to 50$\,a_0$ using the d7Z basis set. 
In each case the results in the neighborhood of $r$ where $B_{n,l}(r)$ changes sign were excluded, reducing the number of points from $D$ to $D'$.
Data points, forming sets of pairs $\{(r_i,B_{n,l}(r_i))\}_{i=1}^{D'}$, were then fitted with analytic functions of the general form
\begin{equation}\label{eq:Bnl}
  \widetilde{B}_{n,l}(r)=-\left(
  r^l\sum_{k=1}^2e^{-a_k r}\sum_{i=1}^4b_{ki}r^{i-1} 
  +r^{n-4}\sum_{\nu\in\mathcal{N}}c_\nu\,f_\nu(\eta r)r^{-\nu}
  \right),
\end{equation}
where $f_\nu(x)$ is the Tang--Toennies damping function defined in Eq.~\eqref{eq:TTdamp} and the values of $\nu\in\mathcal{N}$ used for each function $\widetilde{B}_{n,l}(r)$ are collected in Table~1 in Supplementary Materials~\cite{Supplementary_Materials}.
Any given function has 3 nonlinear parameters ($a_k$ and $\eta$) and $8+|\mathcal{N}|$ linear parameters ($b_{ki}$ and $c_\nu$), and their values were obtained by minimizing the functional
\begin{equation}\label{eq:functional}
\mathcal{F}[\widetilde{B}_{n,l}(r)]
=\sqrt{\frac1{D'}\sum_{i=1}^{D'}\left(\frac{\widetilde{B}_{n,l}(r_i)}{B_{n,l}(r_i)}-1\right)^2}.
\end{equation}

After fixing the $r$-dependent coefficients, the explicit form of the long-range term $V_{\rm L}(R,r,\theta)$ in Eq.~\eqref{eq:Vlms} is directly inferred from Eq.~\eqref{VLR} as
\begin{equation}\label{eq:V_L}
\begin{split}
V_{\rm L}(R,\theta,r)
&=\left(\widetilde{B}_{6,0}(r)\,f_6(q_1 R)\,R^{-6} + \widetilde{B}_{8,0}(r)\,f_8(q_1 R)\,R^{-8}\right) \\ 
&+\left(\widetilde{B}_{7,1}(r)\,f_7(q_1 R)\,R^{-7} + \widetilde{B}_{9,1}(r)\,f_9(q_1 R)\,R^{-9}\right)P_1(\cos\theta) \\
&+\left(\widetilde{B}_{6,2}(r)\,f_6(q_1 R)\,R^{-6} + \widetilde{B}_{8,2}(r)\,f_8(q_1 R)\,R^{-8}\right)P_2(\cos\theta) \\
&+\left(\widetilde{B}_{7,3}(r)\,f_7(q_1 R)\,R^{-7} + \widetilde{B}_{9,3}(r)\,f_9(q_1 R)\,R^{-9}\right)P_3(\cos\theta)\\
&+\widetilde{B}_{8,4}(r)\,f_8(q_1 R)\,R^{-8}\,P_4(\cos\theta) \\
&+\widetilde{B}_{9,5}(r)\,f_9(q_1 R)\,R^{-9}\,P_5(\cos\theta),
\end{split}
\end{equation}
where $q_1$ is a parameter that controls the damping of $V_{\rm L}(R,r,\theta)$ at small $R$.

\subsection{Coordinate transformation for other exotic Helium systems}
When substituting one of the particles $B$ or $C$ in the exotic helium atom, the effective mass ratio $t$ defined in Eq.~\eqref{eq:t} changes to a new value $t'$. Thus, to use the same potential energy surface $V_{\rm fit}(R,r,\theta)$ for systems in which one of the particles is substituted, the following coordinate transformation from $(R',r',\theta')$ to $(R,r,\theta)$ should be performed
\begin{equation}\label{eq:hfi}
\begin{split}
R&=\sqrt{(R')^2+2R'r'(t-t')\cos\theta'+(r')^2(t-t')^2},\\ 
r&=r',\\ 
\cos\theta&=(R'\cos\theta'+r'(t-t'))/R.
\end{split} 
\end{equation}
Once the transformed coordinates are obtained, the interaction energy for the new system follows directly from 
\begin{equation}
V(R',r',\theta')=V_{\rm fit}\big(R(R',r,\theta'),r,\theta(R',r,\theta')\big).
\end{equation}

%%%%%%%%%%%%%%%%%%%%%%%%%%%%%%%%%%%%%%%%%%%%%%%%%%%%%%%%%%%%%%%%%%%%%
%% The appropriate \bibliography command should be placed here.
%% Notice that the class file automatically sets \bibliographystyle
%% and also names the section correctly.
%%%%%%%%%%%%%%%%%%%%%%%%%%%%%%%%%%%%%%%%%%%%%%%%%%%%%%%%%%%%%%%%%%%%%
\bibliography{bibliography}

\end{document}